\documentclass[floatfix,twocolumn,english,aps,arxiv,superscriptaddress]{revtex4-1}
\usepackage{grffile}
\RequirePackage{fix-cm} 
\DeclareMathSizes{10}{10}{5}{5}
\usepackage{gensymb}
\usepackage[T1]{fontenc}
\usepackage[latin9]{inputenc}
\usepackage{color}

\usepackage{babel}
\usepackage{amsmath}
\usepackage{amssymb}
\usepackage{amsfonts}
\usepackage{graphicx}
\usepackage{multirow}
\usepackage{siunitx}
\usepackage{cancel}
\PassOptionsToPackage{normalem}{ulem}
\usepackage{ulem}
\usepackage[version=3]{mhchem}
\usepackage[unicode=true, bookmarks=false, breaklinks=false,pdfborder={0 0 1},colorlinks=false] {hyperref}
\usepackage{subcaption}
\begin{document}

\title{Robust superzone gap opening in incommensurate antiferromagnetic semimetal \ce{EuAg4Sb2} under in-plane magnetic field} 

\author{J. Green}
\affiliation{Department of Physics and Astronomy and California NanoSystems Institute, University of California, Los Angeles, CA 90095, USA}

\author{Arpit Arora}
\affiliation{Division of Physical Sciences, College of Letters and Science, University of California, Los Angeles (UCLA), Los Angeles, CA, USA}
\affiliation{Department of Electrical and Computer Engineering, UCLA, Los Angeles, CA, USA}

\author{Madalynn Marshall}
\affiliation{Neutron Scattering Division, Oak Ridge National Laboratory, Oak Ridge,
Tennessee 37831, USA}

\author{Wanfei Shan}
\affiliation{Division of Physical Sciences, College of Letters and Science, University of California, Los Angeles (UCLA), Los Angeles, CA, USA}

\author{P\'eter Udvarhelyi}
\affiliation{Department of Chemistry and Biochemistry, University of California, Los Angeles, Los Angeles, CA, USA}

\author{Zachary Morgan}
\affiliation{Neutron Scattering Division, Oak Ridge National Laboratory, Oak Ridge,
Tennessee 37831, USA}

\author{Prineha Narang}
\affiliation{Division of Physical Sciences, College of Letters and Science, University of California, Los Angeles (UCLA), Los Angeles, CA, USA}
\affiliation{Department of Electrical and Computer Engineering, UCLA, Los Angeles, CA, USA}

\author{Huibo Cao}
\affiliation{Neutron Scattering Division, Oak Ridge National Laboratory, Oak Ridge,
Tennessee 37831, USA}

\author{Ni Ni}
\email{Corresponding author: nini@physics.ucla.edu}
\affiliation {Department of Physics and Astronomy and California NanoSystems Institute, University of California, Los Angeles, CA 90095, USA}

\begin{abstract}

The interplay between magnetism and charge transport in semimetals has emerged as a fertile ground for discovering novel electronic phenomena. A notable example is the recent discovery of electronic commensuration arising from a spin moir\'e superlattice (SMS), realized as double-$q$ spin modulation in the antiferromagnetic semimetal \ce{EuAg4Sb2}. Here, we investigate the in-plane magnetic-field tunability of the SMS using neutron scattering, magnetic and transport measurements. We reveal an incommensurate noncollinear cycloidal magnetic ground state. Temperature-field phase diagrams constructed with field tilting uncover multiple spin-reoriented phases, suggesting the critical role of in-plane field components in driving magnetic transitions. Despite substantial spin reorientation of the double-$q$ phase, we observe a persistent gap opening, evidenced by strong suppression in both Hall and longitudinal conductivities. Model calculations attribute this robustness to the stability of SMS under tilting fields. Our results establish \ce{EuAg4Sb2} as a tunable platform for exploring spin-texture-driven superzone gap opening in electronic states.

\end{abstract}
\date{\today}
    \maketitle

\textit{Introduction}  The interplay between magnetism, band structure, and charge transport lies at the heart of emergent phenomena in quantum materials. In systems with itinerant electrons and localized magnetic moments, the onset of magnetic order can not only give rise to the anomalous Hall effect or topological Hall effect, but can also dramatically alter the electronic structure, leading to significant changes in transport properties \cite{GdRu2Si2, Gd2PdSi3, EuIn2As2, EuPtSi, GdPtBi, Nd3Al, Nd2Mo2O7, MnSi, MnGe, FeGe, discretized, epitaxialFeGe, MnSi-2, chiralmagnet, THE, breathing, Fe3Sn2, EuAl4, EuGa4, EuGa2Al2}. For instance, in antiferromagnetic semimetals, magnetic ordering often exhibits a larger periodicity than the crystallographic unit cell. This results in a smaller magnetic Brillouin zone (BZ) compared to the crystallographic one, leading to a band folding effect where electronic bands that originally extended over the larger crystallographic BZ are folded into the smaller magnetic BZ. In rare cases, the reconstruction of the BZ can result in the formation of magnetic superzone gaps at the Fermi level \cite{GdSi,GdSi-2, GdSi-3, GdSi-4, DyTe3, CePd5Al2, U2Rh3Si5, NdAlSi, Cr}, causing drastic changes in charge transport measurements.

\begin{figure*}
\centering
    \includegraphics[width=\textwidth]{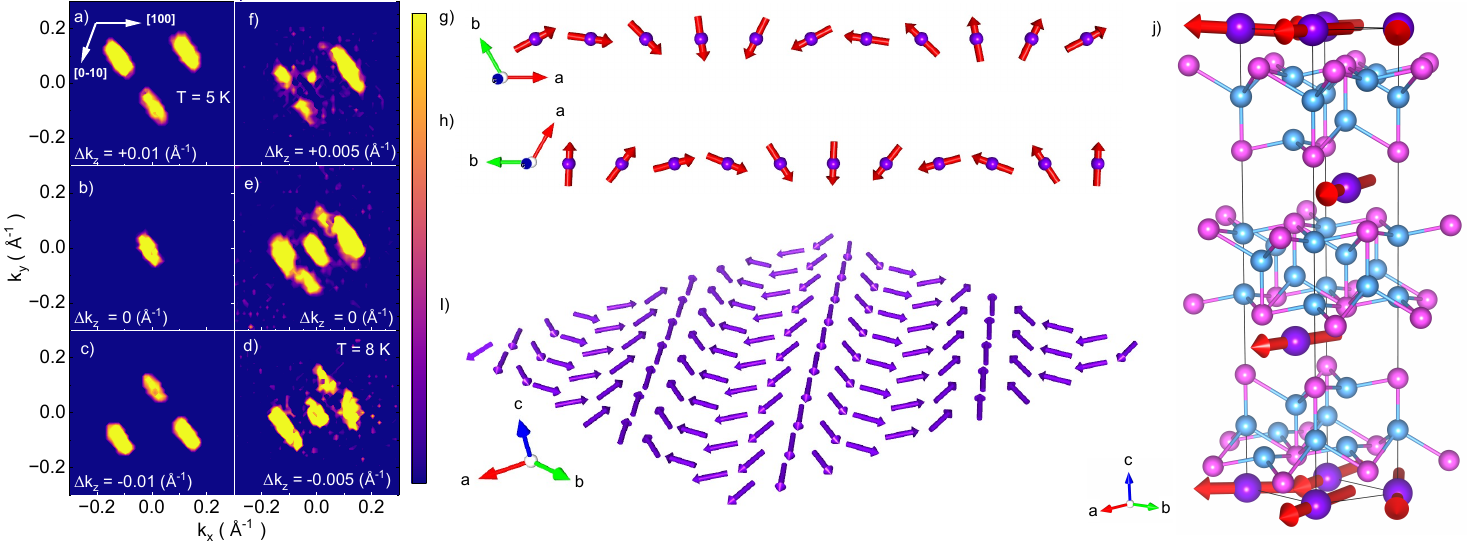}
    \caption{The noncollinear cycloidal magnetic structure of \ce{EuAg4Sb2} at 5 K. Panel (a)-(d) show the single-crystal neutron diffraction of \ce{EuAg4Sb2} around the (101) nuclear peak, where coordinates are given relative to the (101) peak. Panels (a)-(c) are slices taken at $\pm0.01$ \AA$^{-1}$ and $0$ \AA$^{-1}$ at 5 K. Panels (f)-(d) are slices taken at $\pm0.005$ \AA$^{-1}$ and $0$ \AA$^{-1}$ at 8 K. Panel (g)-(i) the elliptical cycloidal spin texture in the $ab$ plane. (j) the moments in one unit cell.}
    \label{MagStruct}
\end{figure*}

Recent studies show that the layered triangular incommensurate antiferromagnetic semimetal family \ce{EuAg4X2} (X = As and Sb) exhibits abnormal magnetotransport properties \cite{Shen-EuAg4As2, Zhu-EuAg4As2, Budko-EuAg4As2, Malick-EuAg4Sb2, Kurumaji-EuAg4Sb2}. Particularly, a sharp drop in longitudinal conductivity $\sigma_{xx}$ and a quench of Hall conductivity $\sigma_{xy}$ were observed in the double-$q$ magnetic phase of \ce{EuAg4Sb2}\cite{Kurumaji-EuAg4Sb2}, suggesting a magnetism-driven gap opening. \ce{EuAg4Sb2} crystallizes in the centrosymmetric space group $R\bar{3}m$ (No.166) and undergoes two weakly first-order magnetic phase transitions at $T_1 =$ 10.6 K and $T_2 =$ 7.5 K at zero field, as evidenced by the observation of sharp lambda-shaped features seen in specific heat measurements \cite{Malick-EuAg4Sb2, Green-EuAg4Sb2}. The electronic structure of EuAg$_4$Sb$_2$ \cite{Green-EuAg4Sb2, Kurumaji-EuAg4Sb2}, revealed through quantum oscillation measurements, angular-resolved photoemission spectroscopy as well as first-principles calculations, consists of three sets of spin-split Fermi pockets: tubular hole pockets and small hourglass-shaped hole pockets at the Brillouin zone center, as well as diamond-shaped electron pockets at the zone boundary. These features suggest a highly anisotropic Fermi surface that is susceptible to interaction with the underlying spin texture. As the magnetic field along $c$ increases, revealed by small angle neutron scattering (SANS), the system undergoes a sequence of incommensurate magnetic phases: ICM1, featuring a single-$q$ modulation, followed by ICM2, a double-$q$ phase, and ICM3, a more complex double-$q$ configuration \cite{Kurumaji-EuAg4Sb2}. The sequence of magnetic phase transitions from ICM3 to ICM2 and finally ICM1 were also realized upon cooling under zero field. Kurumaji et al. pointed out that in the ICM2 phase, since the propagation vector $q_{\rm{ICM2}}$ is close to the diameter $2k_F$ of the hourglass-shaped pockets, the incommensurate double-$q$ magnetic order, likely of the SkL type, introduces a long-wavelength periodic exchange potential. This leads to the electronic commensuration that reconstructs the hourglass pocket via superzone gap formation, drastically influencing transport properties. This effect is analogous to moir\'e-induced minibands in twisted geometric systems, and thus the ICM2 phase is suggested to be a realization of the spin moir\'e superlattice (SMS) \cite{Kurumaji-EuAg4Sb2, Cr}.

Despite this new exciting advancement, key questions remain regarding the robustness of the gap opening in \ce{EuAg4Sb2} and the underlying mechanism of electronic commensuration, both of which are essential for understanding the stability and the role of the proposed SMS. Here, we systematically investigate the evolution of magnetism and charge transport in EuAg$_4$Sb$_2$ as the magnetic field is tilted from the out-of-plane $c$-axis toward the in-plane $a$-axis. Our results reveal a remarkable persistence of the superzone gap opening, as evidenced by a 88\% quenched $\sigma_{xy}$ and an 68\% drop in $\sigma_{xx}$ with the field tilting away from $c$ by 18$^\circ$, despite the spin texture being sensitive to the tilt and undergoing substantial reorientation. This finding indicates that electronic band reconstruction remains stable despite spin reorientation, suggesting the robustness of SMS physics in EuAg$_4$Sb$_2$ and establishing it as a promising platform for exploring magnetically tunable moir\'e phenomena.

 \textit{Results} We first investigated the magnetic structure of the ground state of \ce{EuAg4Sb2}. Figure 1 summarizes the single crystal neutron diffraction measurements taken at DEMAND HB3A at ORNL \cite{HB3A}. At 5 K, in the ICM1 phase, along $k_z$, in addition to the nuclear Bragg peaks, two sets of three magnetic satellite peaks were observed above and below the nuclear (1 0 1) peak, as shown in Fig. 1(a)-(c). The set of three peaks is the result of domain formations in the crystal as they form three $k$-domains with a 120$\degree$ angle between each domain's propagation direction. The magnetic propagation vectors associated with each domain are $\boldsymbol{k_{m}}^{1}$ = (-0.100(1), 0.100(1), 0.04(1)), $\boldsymbol{k_{m}}^{2}$ = (0.100(1), 0, 0.04(1)), and $\boldsymbol{k_{m}}^{3}$ = (0, -0.100(1), 0.04(1)). However, since they are all equivalent, the following discussion will focus on the first domain with the propagation 
 vector $\boldsymbol{k_{m}}^{\textit{ICM1}}$ = (-0.100(1), 0.100(1), 0.04(1)). The small values of the propagation vector components indicate the magnetic periodicity is $10\times$ the crystal unit cell in the $ab$ plane. Data were also collected at $T = 8$ K, in the ICM2 phase, as depicted in panels Fig. 1(d)-(f). The elongation of one of the three satellite peaks suggests peak splitting, consistent with the double-$q$ structure observed in previous SANS experiments \cite{Kurumaji-EuAg4Sb2}. Non-uniform peak intensities of observed peaks are the results of different magnetic structure factors from different $Q$-positions. 
 Table SI summarizes the refined crystal structure from the nuclear Bragg peaks collected at 15 K. Using the magnetic satellite Bragg peaks collected at 5 K, the magnetic structure refinement was made. The goodness of fitting plot is shown in Fig. S1. The magnetic symmetry is $R\Bar{1}$, considering only Eu$^{2+}$ magnetic ions. Figures \ref{MagStruct}(g)-(j) depict the refined magnetic structure. The moment size is refined to be 6.5(2)$\mu_b$ at 5 K. As we can see, in the ICM1 phase, the Eu$^{2+}$ moments lie in the $ab$ plane within the error bar, exhibiting an noncollinear elliptical cycloidal spin texture in the $ab$ plane.

\begin{figure*}
\centering
    \includegraphics[width=\textwidth]{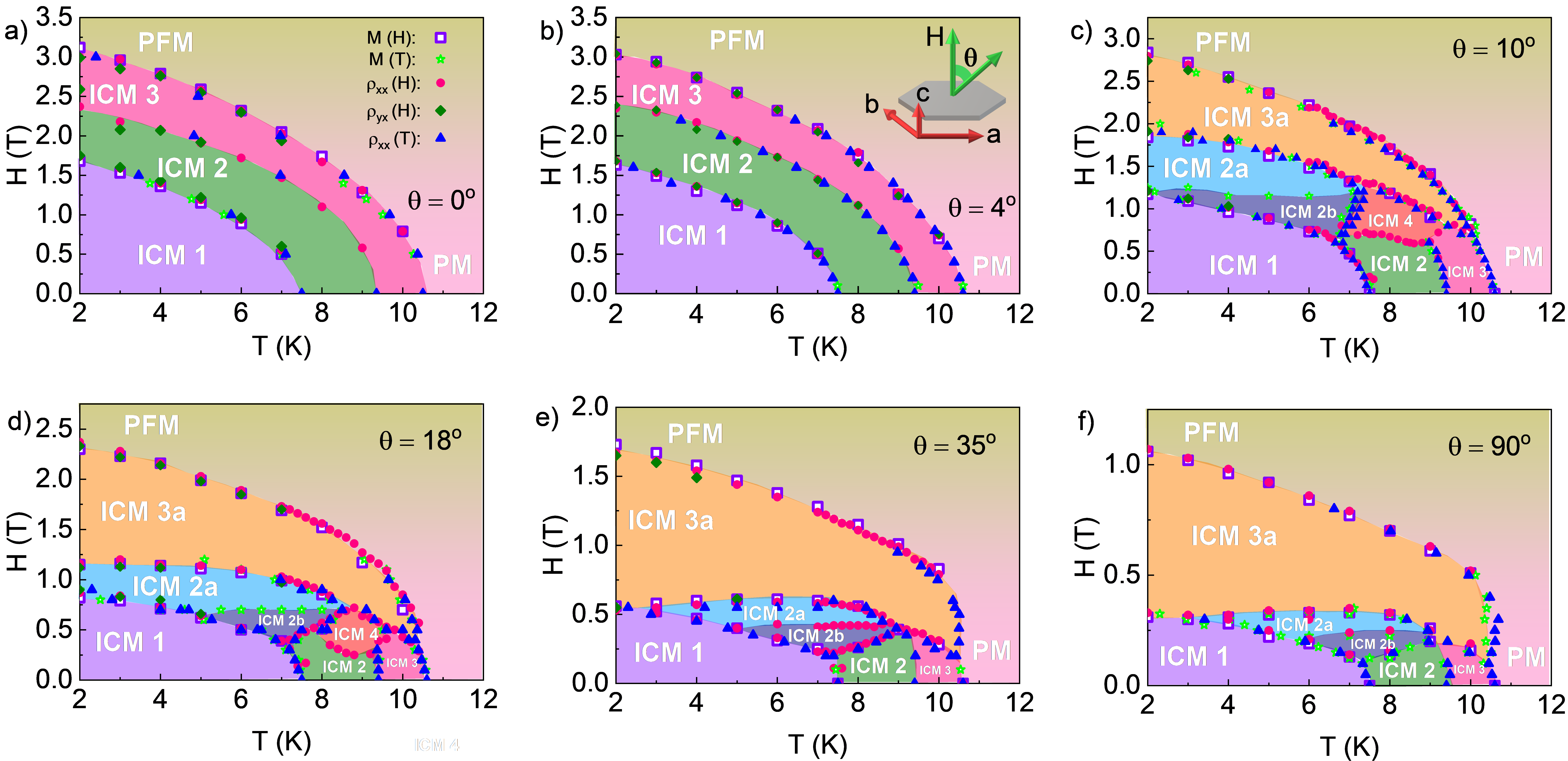}
    \caption{The temperature-field ($T-H$) phase diagram of \ce{EuAg4Sb2} at different tilting angles. They were mapped out based on our comprehensive transport and magnetic property measurements, including isothermal magnetization $M(H)$, magnetic susceptibility $\chi(T)$, temperature-dependent and field-dependent longitudinal resistivity and Hall resistivity, $\rho_{xx}(T)$, $\rho_{xx}(H)$, as well as $\rho_{xy}(H)$. All transport measurements were made with current $I$ applied along the $a$-axis while varying the tilting angle from $0^\circ$ to $90^\circ$. For details, please refer to \cite{SI}.}
    \label{PhaseDiagram}
\end{figure*}

Figure~2 presents the evolution of the temperature-magnetic field ($T\text{-}H$) phase diagrams by rotating the magnetic field from $c$-axis to $a$-axis with the tilting angle $\theta$ at $0^\circ$, $4^\circ$, $10^\circ$, $18^\circ$, $35^\circ$ and $90^\circ$. These phase diagrams, constructed from our magnetic and transport measurements, as shown in Figs. S2-S5 \cite{SI},  reveal rich competing phases. For $\theta = 0^\circ$, at zero field, upon cooling, it transitions from the paramagnetic (PM) phase to the incommensurate double-$q$ noncoplanar ICM3 phase, followed by the incommensurate double-$q$ noncoplanar ICM2 phase, and finally to the incommensurate single-$q$ noncollinear cycloidal ICM1 phase below 7.5 K. As the tilting angle increases to $\theta = 4^\circ$, the overall phase boundaries remain similar to those at $H\parallel c$, indicating that the tilting of the magnetic field has minimal impact on the magnetic phase transitions at this small angle. However, the phase diagram at $\theta = 10^\circ$ shows a drastic change. When the field reaches approximately 0.6 T, with an in-plane component of 0.1 T, spin reorientation emerges, leading to four new magnetic phases, ICM2a, ICM2b, ICM3a, and ICM4, where ICM3a phase was shown to be likely a single-$q$ phase by the SANS taken at $\theta=12$, 2.2 T and 1.5 K \cite{Kurumaji-EuAg4Sb2}. At $\theta = 18^\circ$, all phase boundaries clearly shift to lower fields, resulting in a suppressed ICM2b and ICM4 phase regions. The boundary between the ICM2a and ICM3a phases moves downward more quickly than the others, causing an expansion of the ICM3a phase region while the ICM2a phase region shrinks. At $\theta = 35^\circ$, all phase boundaries continue shifting to lower fields, with the ICM2a/ICM3a boundary moving downward more quickly. The ICM4 phase region becomes indiscernible, while the ICM3a phase continues to expand, suppressing the ICM2a phase to the extent that it is no longer present at 2 K. At $35^\circ\leq\theta \leq 90^\circ$, the overall structure of the phase diagram remains unchanged with all phase boundaries continually shifting to lower fields.

Figure~3 summarizes the $\sigma_{xx}(H)$ and $\sigma_{xy}(H)$, with the tilting angle $\theta$ varied from $0^\circ$ to $90^\circ$ and current $I$ directed along the $a$-axis. Electrical transport is strongly coupled to the magnetism. At $H \parallel c$ and 2 K, upon transitioning from the ICM1 phase to the ICM2 phase, $\sigma_{xx}(H)$ exhibits a sharp decrease by 50\%, while $\sigma_{xy}$ abruptly drops by 96\%, being the characteristic transport signatures of the opening of the superzone gap due to SMS formation in \ce{EuAg4Sb2}. At $\theta = 4^\circ$, the $T-H$ phase diagram remains quite similar to that at $H//c$. Meanwhile, compared to the electrical transport at $H\parallel c$, both $\sigma_{xx}$ and $\sigma_{xy}$ exhibit insignificant changes, with electrical transport showing a smooth transition across the phase boundaries between ICM2 and ICM3. At $\theta = 10^\circ$ and $18^\circ$, where the ground state transitions from ICM1 to ICM2a and subsequently to ICM3a, a new envelope of Hall conductivities emerges. Firstly, upon entering ICM2a from ICM1, sharp drops by 92\% in $\sigma_{xy}$ and $\sim$ 65\% in $\sigma_{xy}$ are observed. keep in mind that the drops are 96\% and 50\%, respectively when going from ICM1 to ICM2 at $\theta = 0$. Secondly, two plateau features appear, with a step at the phase boundary between ICM2a and ICM3a in both $\sigma_{xx}$ and $\sigma_{xy}$. Specifically, $\sigma_{xx}$ decreases further in the ICM2a phase compared to the ICM2 phase, while the decrease is much smaller in the ICM3a phase, giving rise to the two-plateau feature. The emergence of the two-plateau feature clearly reflects the distinct nature of the transitions from ICM2 to ICM2a and from ICM3 to ICM3a. At $\theta = 35^\circ$, the ground state transitions directly from ICM1 to ICM3a without passing through ICM2a. Consequently, the transport signatures of the ICM2a phase vanish and the two-plateau feature disappears, while the drop in Hall conductivities when transitioning into ICM3a remains, leading to a one-step feature. 

\textit{Discussion} As a semimetal with triangularly-arranged Heisenberg Eu$^{2+}$ spins and highly anisotropic Fermi surfaces, various energy scales compete, including dominant Ruderman-Kittel-Kasuya-Yosida (RKKY) interaction, easy-plane anisotropy, geometric frustrations, and Zeeman energies, leading to the rich incommensurate magnetic states in \ce{EuAg4Sb2}. When the field tilts from $c$ to $a$, the in-plane field component grows and dominantly controls the evolution of the phase diagram. A clear example is the phase boundary between the ICM2a and ICM2b phases. This boundary appears at 1.2 T, 0.7 T, 0.4 T, and 0.23 T for tilt angles of 10$^\circ$, 18$^\circ$, 35$^\circ$, and 90$^\circ$, respectively. Notably, the corresponding in-plane field components at these critical points are remarkably similar, falling within a narrow range of 0.21 T to 0.23 T. A similar trend is observed for the ICM3a-ICM2a phase boundary, where the in-plane components of the critical fields lie between 0.32 T and 0.35 T. These observations suggest that the in-plane field component plays a key role in driving these phase transitions. 

Figures 3(c) and 3(d) summarize the drop percentage in $\sigma_{xy}$ and $\sigma_{xx}$ when transitioning from the ICM1 phase to the ICM2/ICM2a phase, as well as the increase percentage observed during the transition from the ICM3a phase to the PFM phase, all plotted as a function of tilting angle. For reference, the temperature dependence of the drop percentages at ICM2 phase with $\theta=0$ are shown in the insets. Remarkably, the sharp drop in both $\sigma_{xy}$ and $\sigma_{xx}$ persists across both the ICM2 and ICM2a phases. Even at $\theta = 18^\circ$ and 2 K, within the ICM2a phase, the suppression in $\sigma_{xy}$ remains comparable to that observed in the ICM2 phase at $\theta = 0^\circ$ and 3 K, while the decrease in $\sigma_{xx}$ exceeds that in the ICM2 phase by nearly 15\%. These observations suggest that the gap opening in the ICM2a phase remains substantial, likely comparable to or even exceeding that in the ICM2 phase. In contrast, the changes in $\sigma_{xy}$ and $\sigma_{xx}$ in the ICM3a phase are significantly smaller, being around 75\% and 35\% at 35$^\circ$ and 2 K, respectively, indicating a much more suppressed gap opening.

\begin{figure}
\centering
\includegraphics[width=3.5in]{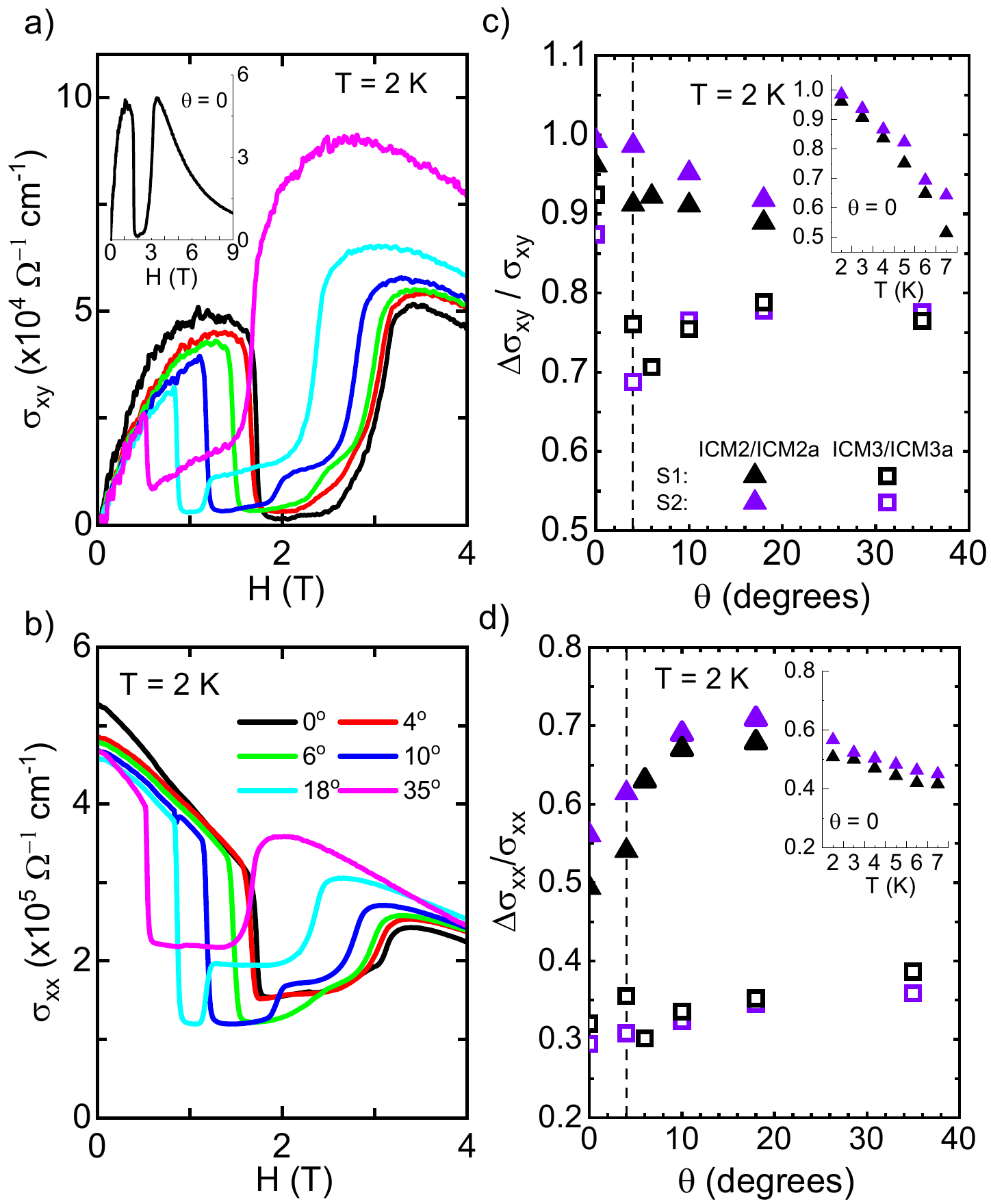}
    \caption{(a-b) $\sigma_{xy}$ (a) and $\sigma_{xx}$ (b) of sample S1 at tilt angles 0$^\circ$, 4$^\circ$, 6$^\circ$, 10$^\circ$, 18$^\circ$, and 35$^\circ$. The inset illustrates the $\sigma_{xy}$ at 0$^\circ$ extended for fields up to 9 T. (c-d) The drop percentage observed in $\sigma_{xy}$ (c) and $\sigma_{xx}$ (d) vs. tilt angle of Sample S1 and S2, where Sample S2 has a larger geometric mean of mobility, as showin in Fig. S6. Left of the dash line: from ICM1 to ICM2, or from ICM3 to PFM. Right to the dash line: from ICM1 to ICM2a, or from ICM3a to PFM. For details, see SI\cite{SI}. Inset: the temperature dependence of the drop percentage for 0$^\circ$. }
    \label{Transport-Angle}
\end{figure}

\begin{figure*}
\centerline{\includegraphics[width=\textwidth]{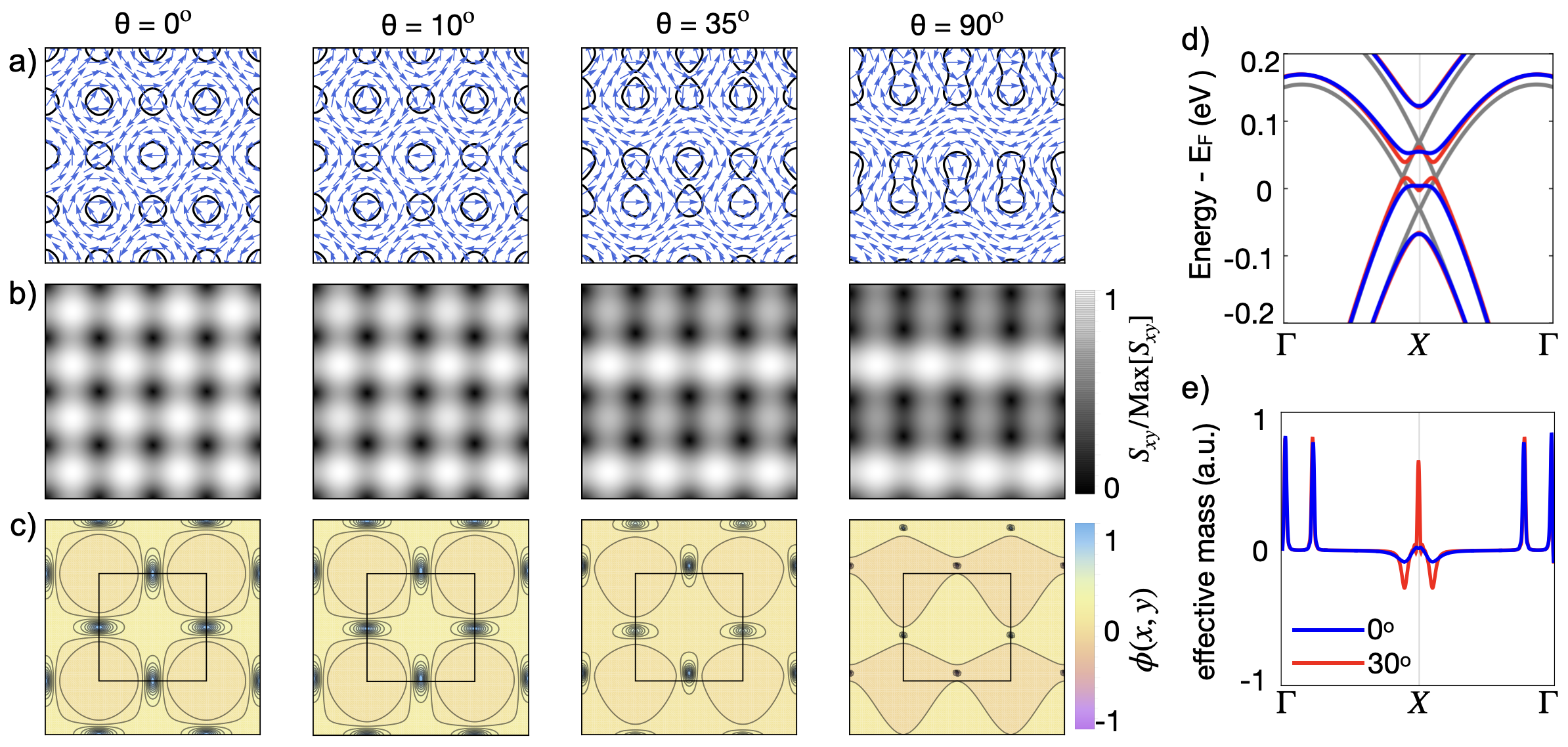}}
     \caption{Square skyrmion lattice obtained with double-q spin modulation for $S_0= -0.5$, and its evolution with tiltig magnetic field away from $c$ axis. With increasing $\theta$, the spins cant along the in-plane component, leading to a skewed (a) skyrmion core, which leads to (b) diffused interference of the spin texture weakening the moir\'e superlattice potential. For a field along the $a$ axis, the SkL transforms to a striped pattern. (c) Skyrmion density corresponding to panels (b) and (c). (d) Electronic band structure along high symmetry points in the magnetic BZ and band folding due to SMS for $S_0=-0.3$; gray line denotes the ferromagnetic phase. (e) Effective mass of the SMS folded bands around the Fermi surface.}
    \label{sms}
\end{figure*}

Given that the gap opening is robust and comparable in both the ICM2 and ICM2a phases, it is reasonable to suggest that the SMS extends into the ICM2a phase. This raises the question of how the SMS remains stable despite spin reorientation. To elucidate this, we analyze the evolution of spin texture and the corresponding band structure under the tilting of the magnetic field. We model the spin texture of SkL \cite{Kurumaji-EuAg4Sb2} with double-$q$ spin modulation $\mathbf{S}(x,y) = [\hat{\mathbf{x}} \sin(qy) + \hat{\mathbf{y}} \sin(qx) + \hat{\mathbf{z}}\{\cos(qx) +\cos(qy)\}]/2 + \mathbf{S}_0$ where $\{S_x,S_y,S_z\}$ describe the intrinsic spin texture of the magnetic lattice and $\mathbf{S}_0=S_0\{\sin\theta, 0 ,\cos\theta\}$ captures the effect of the magnetization due to external magnetic field. This double-$q$ modulation of spins forms a square magnetic superlattice,  enforcing a spin-induced moir\'e potential on the electronic system, resulting in the SMS. 

With fixed $S_0 = -0.5$ we track the in-plane spin textures $\{S_x,S_y\}$, their magnitude, and skyrmion density. As shown in Fig.~\ref{sms}(a), as $\theta$ increases the spins align with the in-plane component of the field, leading to skewed skyrmion cores. This, in turn, diffuses the interference of the spin pattern, noted by $S_{xy} = \sqrt{S_x^2 + S_y^2}$, manifesting in weakening of moir\'e potential, see Fig.~\ref{sms}(b). For $\theta = 90^\circ$, the SkL is washed away, and a striped pattern emerges. The corresponding changes in skyrmion density are shown in Fig.~\ref{sms}(c). The control of SMS with an in-plane magnetic field and the appearance of a striped pattern are reminiscent of strain effects on geometric moir\'e properties~\cite{Sinner2023Strain}. This may suggest the in-plane field can offer a potential knob for engineering nonreciprocal responses in emergent SMS materials.

To get a more complete understanding, we next examine the emergent SMS electronic structure under a tilted magnetic field. We construct an effective model for simulating the low-energy minibands around the Fermi level with a double-exchange model on a triangular lattice interfering with a magnetic square superlattice~\cite{Kurumaji-EuAg4Sb2,nagaosaSkL}, see SI. We consider exchange interaction to be near resonant with the Fermi level, a necessary condition to observe SMS~\cite{Kurumaji-EuAg4Sb2}. The key factor in tracking the sharp changes in $\sigma_{xx}$ and $\sigma_{xy}$ is the gapping of the electronic bands around the Fermi level. This is possible with a stable SkL wherein the combined effect of SMS and an out-of-plane magnetic field generates non-trivial winding, enabling interaction between SMS and spins of the carriers, leading to effects beyond spin splitting. As shown in Fig.~\ref{sms}(d), the SMS leads to band reconstruction in the magnetic BZ, thus capable of modifying the electron transport. Additionally, for SkL most pronounced at $\theta = 0$, the electronic bands gap out, leading to strong renormalization of the effective mass, which in turn affects the mobility, and thus electronic transport. For $\theta \neq 0$, the gap around the Fermi level comes down, increasing the effective mass and mobility, as shown in Fig.~\ref{sms}(e).

We note that although this simple single-layer double-$q$ spin picture presented above successfully predicts the robustness of the superzone gap opening and the SMS with tilting angles, the predicted suppression of gap opening and enhancement of carrier mobility with increasing tilt angle does not align well with the experimentally observed stronger suppression of $\sigma_{xx}$ in the ICM2a phase. This discrepancy may be attributed to small but finite spin modulations along the $c$-axis, which could hybridize the electronic spins with the underlying magnetic texture and help sustain the gap even under in-plane fields. Furthermore, despite the changes in $\sigma_{xx}$ being much smaller in the ICM3a phase, the plateau feature still indicates the presence of partial gap openings there, possibly suppressed but not absent. Since the ICM3a phase is likely a single-$q$ phase identified by the SANS taken at $\theta=12$, 2.2 T and 1.5 K \cite{Kurumaji-EuAg4Sb2}, the simple single-layer double-$q$ spin picture presented above does not account for this behavior. Such an observation may be understood within the density-wave framework. For example, in GdSi \cite{GdSi}, nested itinerant electrons mediate RKKY interactions that align local moments along a propagation vector $q$ that matches the nesting vector, and in turn, the resulting single-$q$ incommensurate magnetic order gaps states at the nested regions of the Fermi surface. Further investigation using complementary experimental and theoretical techniques, including scanning tunneling microscopy and supercell DFT calculations, are urged to clarify the stronger SMS-induced gap opening in the ICM2a phase, to elucidate the Fermi surface nesting condition and gap-opening mechanism in the ICM3a phase, and to understand the role of the out-of-plane spin texture in enabling gap formation in ICM3a but not in ICM1, despite both being single-$q$ incommensurate states with similar $q$ vectors.

\textit{Conclusion} We reveal an incommensurate noncollinear cycloidal magnetic ground state of \ce{EuAg4Sb2}. When the magnetic field is tilted from the $c$ axis toward the $a$ axis, the in-plane field components drive the system into spin-reoriented phases, for instance, transitioning the double-$q$ ICM2 phase into the spin-reoriented ICM2a phase, and the double-$q$ ICM3 phase into the spin reoriented single-$q$ ICM3a phase. Despite substantial spin reorientation in the ICM2a phase, we observe stronger suppression of longitudinal conductivity and slightly reduced suppression of Hall conductivity with increasing tilt angle, compared to the SMS-hosting ICM2 phase. This behavior suggests a robust superzone gap opening and persistent SMS. This is consistent with our model calculation, where we show that the SMS arising from the double-$q$ SkL modulation remains stable under titling fields, which leads to folded electronic bands and gap openings around the Fermi level. Furthermore, while comparing the charge transport in ICM3 and ICM3a phases, the drop percentage in longitudinal conductivity remains comparable while the drop percentage in Hall conductivity becomes smaller, suggesting a reduced extent of gap opening in ICM3a. Our work demonstrates that  \ce{EuAg4Sb2} serves as a versatile platform for investigating spin-texture-induced superzone gap formation and its impact on electronic transport.

\section*{Acknowledgments}
NN thanks the useful discussion with Prof. Igor I. Mazin and Prof. Takashi Kurumaji. Experimental work at UCLA was supported by the U.S. Department of Energy (DOE), Office of Science, Office of Basic Energy Sciences under Award Number DE-SC0021117. M. M and H. C. acknowledge the support from U.S. DOE BES Early Career Award KC0402010 under Contract No. DE-AC05-00OR22725 and the U.S. DOE, Office of Science User Facility operated by the ORNL. 
A. A., W. S., P. U. and P. N. acknowledge the support by Quantum Science Center, a National Quantum Information Science Center of the U. S. Department of Energy, the Office of Science, Basic Energy Sciences, Materials Sciences and Engineering Division of the U. S. Department of Energy, Gordon and Betty Moore Foundation Grant No. 8048, and the John Simon Guggenheim Memorial Foundation (Guggenheim Fellowship).

\medskip

\bibliographystyle{apsrev4-1}
\bibliography{SbMBT}

\begin{thebibliography}{}

\bibitem{GdRu2Si2} N. D. Khanh et al., Nanometric square skyrmion lattice in a centrosymmetric tetragonal magnet, Nat. Nanotechnol. 15, 444 (2020).

\bibitem{Gd2PdSi3} T. Kurumaji, T. Nakajima, M. Hirschberger, A. Kikkawa, Y. Yamasaki, H. Sagayama, H. Nakao, Y. Taguchi, T. Arima, and Y. Tokura, Skyrmion lattice with a giant topological Hall effect in a frustrated triangular-lattice magnet, Science 365, 914 (2019).

\bibitem{EuIn2As2} J. Yan, Z. Z. Jiang, R. C. Xiao, W. J. Lu, W. H. Song, X. B. Zhu, X. Luo, Y. P. Sun, and M. Yamashita, Field-induced topological Hall effect in antiferromagnetic axion insulator candidate ${\mathrm{EuIn}}_{2}{\mathrm{As}}_{2}$, Phys. Rev. Res. 4, 013163 (2022).

\bibitem{EuPtSi} M. Kakihana et al., Giant Hall Resistivity and Magnetoresistance in Cubic Chiral Antiferromagnet EuPtSi, J. Phys. Soc. Jpn. 87, 023701 (2018).

\bibitem{GdPtBi} T. Suzuki, R. Chisnell, A. Devarakonda, Y.-T. Liu, W. Feng, D. Xiao, J. W. Lynn, and J. G. Checkelsky, Large anomalous Hall effect in a half-Heusler antiferromagnet, Nature Phys 12, 1119 (2016).

\bibitem{Nd3Al} D. Singh, J. Nag, S. Yadam, V. Ganesan, A. Alam, and K. G. Suresh, Colossal anomalous Hall conductivity and topological Hall effect in ferromagnetic kagome metal Nd3Al, Applied Physics Letters 123, 171902 (2023).

\bibitem{Nd2Mo2O7} Y. Taguchi, Y. Oohara, H. Yoshizawa, N. Nagaosa, and Y. Tokura, Spin Chirality, Berry Phase, and Anomalous Hall Effect in a Frustrated Ferromagnet, Science 291, 2573 (2001).

\bibitem{MnSi} A. Neubauer, C. Pfleiderer, B. Binz, A. Rosch, R. Ritz, P. G. Niklowitz, and P. B\"{o}ni, Topological Hall Effect in the A Phase of MnSi, Phys. Rev. Lett. 102, 186602 (2009).

\bibitem{MnGe} N. Kanazawa, Y. Onose, T. Arima, D. Okuyama, K. Ohoyama, S. Wakimoto, K. Kakurai, S. Ishiwata, and Y. Tokura, Large Topological Hall Effect in a Short-Period Helimagnet MnGe, Phys. Rev. Lett. 106, 156603 (2011).

\bibitem{FeGe} J. C. Gallagher, K. Y. Meng, J. T. Brangham, H. L. Wang, B. D. Esser, D. W. McComb, and F. Y. Yang, Robust Zero-Field Skyrmion Formation in FeGe Epitaxial Thin Films, Phys. Rev. Lett. 118, 027201 (2017).

\bibitem{discretized} N. Kanazawa, M. Kubota, A. Tsukazaki, Y. Kozuka, K. S. Takahashi, M. Kawasaki, M. Ichikawa, F. Kagawa, and Y. Tokura, Discretized topological Hall effect emerging from skyrmions in constricted geometry, Phys. Rev. B 91, 041122 (2015).

\bibitem{epitaxialFeGe} S. X. Huang and C. L. Chien, Extended Skyrmion Phase in Epitaxial $\mathrm{FeGe}(111)$ Thin Films, Phys. Rev. Lett. 108, 267201 (2012).

\bibitem{MnSi-2} Y. Li, N. Kanazawa, X. Z. Yu, A. Tsukazaki, M. Kawasaki, M. Ichikawa, X. F. Jin, F. Kagawa, and Y. Tokura, Robust Formation of Skyrmions and Topological Hall Effect Anomaly in Epitaxial Thin Films of MnSi, Phys. Rev. Lett. 110, 117202 (2013).

\bibitem{chiralmagnet} T. Schulz, R. Ritz, A. Bauer, M. Halder, M. Wagner, C. Franz, C. Pfleiderer, K. Everschor, M. Garst, and A. Rosch, Emergent electrodynamics of skyrmions in a chiral magnet, Nature Phys 8, 301 (2012).

\bibitem{THE} N. Nagaosa and Y. Tokura, Topological properties and dynamics of magnetic skyrmions, Nature Nanotech 8, 899 (2013).

\bibitem{breathing} M. Hirschberger et al., Skyrmion phase and competing magnetic orders on a breathing kagomé lattice, Nat Commun 10, 5831 (2019).

\bibitem{Fe3Sn2} H. Li, B. Ding, J. Chen, Z. Li, Z. Hou, E. Liu, H. Zhang, X. Xi, G. Wu, and W. Wang, Large topological Hall effect in a geometrically frustrated kagome magnet Fe3Sn2, Applied Physics Letters 114, 192408 (2019).

\bibitem{EuAl4} T. Shang, Y. Xu, D. J. Gawryluk, J. Z. Ma, T. Shiroka, M. Shi, and E. Pomjakushina, Anomalous Hall resistivity and possible topological Hall effect in the ${\mathrm{EuAl}}_{4}$ antiferromagnet, Phys. Rev. B 103, L020405 (2021).

\bibitem{EuGa4}  S. Lei et al., Weyl nodal ring states and Landau quantization with very large magnetoresistance in square-net magnet EuGa4, Nat Commun 14, 5812 (2023).

\bibitem{EuGa2Al2} J. M. Moya et al., Incommensurate magnetic orders and topological Hall effect in the square-net centrosymmetric ${\mathrm{EuGa}}_{2}{\mathrm{Al}}_{2}$ system, Phys. Rev. Mater. 6, 074201 (2022).

\bibitem{GdSi} Y. Feng et al., Incommensurate antiferromagnetism in a pure spin system via cooperative organization of local and itinerant moments, Proceedings of the National Academy of Sciences 110, 3287 (2013).

\bibitem{GdSi-2} Y. Feng, D. M. Silevitch, J. Wang, A. Palmer, N. Woo, J.-Q. Yan, Z. Islam, A. V. Suslov, P. B. Littlewood, and T. F. Rosenbaum, Evolution of incommensurate spin order with magnetic field and temperature in the itinerant antiferromagnet GdSi, Phys. Rev. B 88, 134404 (2013).

\bibitem{GdSi-3} Y. Feng, J. Wang, A. Palmer, J. A. Aguiar, B. Mihaila, J.-Q. Yan, P. B. Littlewood, and T. F. Rosenbaum, Hidden one-dimensional spin modulation in a three-dimensional metal, Nat Commun 5, 4218 (2014).

\bibitem{GdSi-4} Y. Feng et al., Linear magnetoresistance in the low-field limit in density-wave materials, Proceedings of the National Academy of Sciences 116, 11201 (2019).


\bibitem{DyTe3} N. Ru, J.-H. Chu, and I. R. Fisher, Magnetic properties of the charge density wave compounds $R{\text{Te}}_{3}$ ($R=\text{Y}$, La, Ce, Pr, Nd, Sm, Gd, Tb, Dy, Ho, Er, and Tm), Phys. Rev. B 78, 012410 (2008).

\bibitem{CePd5Al2} T. Onimaru et al., Giant Uniaxial Anisotropy in the Magnetic and Transport Properties of CePd5Al2, J. Phys. Soc. Jpn. 77, 074708 (2008).

\bibitem{U2Rh3Si5} B. Becker, S. Ramakrishnan, A. A. Menovsky, G. J. Nieuwenhuys, and J. A. Mydosh, Unusual Ordering Behavior in Single-Crystal ${U}_{2}{\mathrm{Rh}}_{3}{\mathrm{Si}}_{5}$, Phys. Rev. Lett. 78, 1347 (1997).

\bibitem{NdAlSi} J. Gaudet et al., Weyl-mediated helical magnetism in NdAlSi, Nat. Mater. 20, 1650 (2021).

\bibitem{Cr} Y. Feng, Y. Wang, T. F. Rosenbaum, P. B. Littlewood, and H. Chen, Quantum interference in superposed lattices, Proceedings of the National Academy of Sciences 121, e2315787121 (2024).

\bibitem{Shen-EuAg4As2} B. Shen, C. Hu, H. Cao, X. Gui, E. Emmanouilidou, W. Xie, and N. Ni, Structural distortion and incommensurate noncollinear magnetism in EuAg 4 As 2, Phys. Rev. Materials 4, 064419 (2020).

\bibitem{Zhu-EuAg4As2} Q. Zhu, L. Li, Z. Yang, Z. Lou, J. Du, J. Yang, B. Chen, H. Wang, and M. Fang, Metamagnetic transitions and anomalous magnetoresistance in EuAg4As2 crystals, Sci. China Phys. Mech. Astron. 64, 227011 (2020).

\bibitem{Budko-EuAg4As2} S. L. Bud'ko, L. Xiang, C. Hu, B. Shen, N. Ni, and P. C. Canfield, Pressure tuning of structural and magnetic transitions in \ce{EuAg4As2}, Phys. Rev. B 101, 195112 (2020).

\bibitem{Malick-EuAg4Sb2} S. Malick, H. \'Swi\k{a}tek, J. B\l awat, J. Singleton, and T. Klimczuk, Large magnetoresistance and first-order phase transition in antiferromagnetic single-crystalline \ce{EuAg4Sb2}, Phys. Rev. B 110, 165149 (2024).

\bibitem{Kurumaji-EuAg4Sb2} T. Kurumaji et al., Electronic commensuration of a spin moir\'e superlattice in a layered magnetic semimetal, Science Advances 11, eadu6686 (2025).

\bibitem{Green-EuAg4Sb2} J. Green et al., Mapping the three-dimensional fermiology of the triangular lattice magnet ${\mathrm{EuAg}}_{4}{\mathrm{Sb}}_{2}$, Phys. Rev. B 111, 085139 (2025).

\bibitem{HB3A} B. C. Chakoumakos, H. Cao, F. Ye, A. D. Stoica, M. Popovici, M. Sundaram, W. Zhou, J. S. Hicks, G. W. Lynn, and R. A. Riedel, J. Appl. Crystallogr. 44, 655 (2011).

\bibitem{SI} Supplementary Information: Additional experimental and theoretical details, including data used to construct $T-H$ diagrams and some photos of experimental setup.

\bibitem{Sinner2023Strain} A. Sinner, P. A. Pantele\'on and F. Guinea, Strain-induced quasi-1D channels in twisted moir\'e lattices, Phys. Rev. Lett. 131, 166402 (2023).

\bibitem{nagaosaSkL}  K. Hamamoto, M. Ezawa, and N. Nagaosa, Quantized topological Hall effect in skyrmion crystal, Phys. Rev. B 92, 115417 (2015).


\end{thebibliography}

\end{document}